\documentclass[fleqn,twoside]{article}
\usepackage{espcrc2}

\usepackage{graphicx}
\usepackage[figuresright]{rotating}

\def\FIG #1 #2 [#3] #4\par{%
  \begin{figure}[!h] \begin{center}%
    \includegraphics*[#3]{#2}%
    \\ \parbox{10cm}{%
    \caption{\label{#1}#4}}%
  \end{center}\end{figure}%
}

\def\FIGG #1 #2 #3 [#4] #5\par{%
  \begin{figure}[!h]
    \includegraphics*[#4]{#2}
    \hfill
    \includegraphics*[#4]{#3}
    \caption{\label{#1}#5}
  \end{figure}
}

\newcommand{\AmS}{{\protect\the\textfont2
  A\kern-.1667em\lower.5ex\hbox{M}\kern-.125emS}}

\title{Time-dependent thermal effects in GRB afterglows}

\author{
K.A.Postnov\address[SAI]{Sternberg Astronomical Institute, 
119992 Moscow,  Russia}
\thanks{The work is partially supported through RFBR grant 02-02-16500},
S.I.Blinnikov\address{Institute of Theoretical and Experimental Physics,
117218 %sb
Moscow, Russia},        
D.I.Kosenko\addressmark[SAI],
E.I.Sorokina\addressmark[SAI]
}
\begin{document}

\begin{abstract}
Time-dependent thermal effects should accompany standard
non-thermal afterglows of GRB when gamma-rays pass 
through inhomogeneous surroundings of the GRB site. 
Thermal relaxation of an optically thin plasma is calculated 
using time-dependent collisional ionization of the 
plasma ion species. X-ray emission lines are similar 
to those found in the fading X-ray afterglow of GRB 011211. 
Thermal relaxation of 
clouds or shells around the GRB site could also contribute
to the varying late optical GRB afterglows, such as in 
GRB021004 and GRB030329.      
\vspace{1pc}
\end{abstract}

\maketitle

\section{Time-dependent X-ray emission from a thermal plasma}

Determination of physical parameters of astrophysical sources with plasma
densities up to $10^{11}-10^{12}$ cm$^{-3}$ should include time-dependent
ionization processes. The cooling time for a completely
ionized plasma composed of electrons and ions with charge $Z$ 
(due to free-free radiation alone; allowing for the recombination 
emission decreases this time severalfold) is 
$
t_c\approx 2\times 10^{15}[s] n_e \sqrt{T/10^8}/Z 
$
%This is an upper limit for the real cooling time.
In real GRB, the duration of gamma-ray emission is 1-100 s
and can be much smaller than the plasma cooling time. 

%\FIGG cool eloses elosesme [width=.45\textwidth] Plasma
%cooling function $\varepsilon^{-}$ (erg cm$^{3}$ s$^{-1}$). {\bf
%Left panel}: solar composition, then the emission measure for
%$T=10^8$ K is $\mathrm{EM} = n^2\,V = L_X/\varepsilon^{-}\simeq
%3.7\times10^{68}\t{cm}^3$; {\bf Right panel}: metal-rich medium
%(without H and He), $\mathrm{EM} = n^2\,V = L_X/\varepsilon^{-}
%\simeq 1.4\times10^{65}\t{cm}^3$

% Fig. Cooling rates for different plasmas

{\it X-ray lines in the GRB 011211 afterglow}. XMM-Newton observations of GRB011211 afterglow
\cite{Reeves_et_al_2002} 11 hrs after
the GRB are best fit by a thermal plasma 
continuum with $T$ of several (2-4) keV
and emission lines (identified with highly ionized metal species, 
except for iron)
faded away over $\sim 10^4$ s. 
In contrast to purely geometrical
interpretation of these observations (which requires a very high 
density shell with $n_e\sim 10^{15}$ cm), our time-dependent approach 
identifies this time with {\it thermal relaxation time} of
a plasma heated by passing gamma-rays up to a temperature of several keV     
(deposited energy $\epsilon_0\sim 10$ keV per nucleon). 
This time implies plasma density $n_e\sim 10^{11}-10^{12}$ cm$^{-3}$.
Then the  
observed X-ray emission measure $EM=n_e^2V\sim 10^{69}$ cm$^{-3}$ % sb normally EM
means $V\sim 10^{47}-10^{45}$ cm$^{3}$. 

{\it Need for clouds}. The optical thickness
w.r.t. Thompson scattering $n\sigma_T < 1$ requires 
a {\it clumpy surroundings} 
with the number of clouds $N_{cl}\sim 10^6$ of size 
$l_{cl}\sim 10^{13}$ cm and the total volume providing the required emission
measure. (This is an {\it upper limit}: the  realistic cooling function
yields $EM\sim 4 \times 10^{68}$ cm$^{-3}$ for normal cosmic abundance and
$EM\sim 10^{65}$ cm$^{-3}$ for plasma without H and He and
other elements in solar ratio). 
The ISM with such cloudy properties is common around e.g. late AGB stars
\cite{Bains_et_al_2003}.
Sub-AU clouds of the density $\sim 10^{12}$ cm$^{-3}$ 
were also involved by \cite{Dermer_and_Mitman_1999} to explain the observed short
time-scale variability on GRB light curves in the external shock scenario.

{\it The energy deposit into surroundings}. 
The required energy deposit per nucleon $\epsilon_0\sim$ 10 keV 
limits the external space where such a medium can be heated by gamma rays
to the distance $
d=\sqrt{\frac{\sigma_\gamma E_\gamma}{4\pi\epsilon_0}}\sim 2\times
10^{17}\hbox{cm}$ 
where $E_\gamma=5\times 10^{52}$ erg is the total isotropic GRB energy, 
$\sigma_\gamma\approx 0.1 \sigma_T$ is the effective gamma-ray deposition
cross section for heating the environment per nucleon 
\cite{Ambwani_and_Sutherland_1988}. This distance is independent of the 
assumed isotropy/beaming of the GRB emission. Again, this distance
is strikingly similar to what is needed in the external shock scenario
\cite{Dermer_and_Mitman_1999}. 

{\it GRB jet opening angle} $\theta$ should be specified to calculate the
total volume of the circumburst matter illuminated by GRB.
Assuming time delay $t_l=10^4$ s to be entirely due to thermal relaxation 
and not to geometrical retardation yields the relation
$d(1-\cos\theta)=(ct_l)/(1+z)\simeq 10^{14}\hbox{cm}$  
For small $\theta$, $d=2\times 10^{17}$ cm and $z=2.14$ we get 
$\theta\sim \sqrt{2ct_l/(1+z)d}\sim 0.05$. Such jet collimation angles 
$2\theta\sim 0.1$ are common in  GRB \cite{Frail_et_al_2000}. 
The total illuminated 
volume is thus $V\sim \theta^2 d^3\approx 3\times 10^{49}$ cm$^{-3}$
which implies the cloud filling factor $f\sim 10^{45}/10^{49}\sim 10^{-4}$. 
So the physically acceptable structure of clumpy ISM can 
actually be the source of the observed thermal X-rays from GRB 011211. 
For example, such a medium can be produced if several hundred late-type
stars with strong winds fall within the GRB cone. Unusual in our
Galaxy, this could possibly take place in star-forming 
galaxies at large redshifts where GRB are observed.

{\it Modeling the X-ray spectrum}.
We use time-dependent code to compute the state of 
ionization and radiation of thermal plasma 
which include the basic elementary processes (collisional 
ionization, autoionization, photorecombination, dielectronic recombination,
ion charge exchange) and computes free-free, free-bound,
bound-bound and  two-photon emission for ions of all types. 
The parameters include time-dependent temperature $T$, 
density $\rho$ and chemical composition. 
%The cooling of optically thin plasma is determined
%from the equation
%$$
%\frac{dT}{dt}=-\frac{\epsilon^-(T)n_e^2}{C_V\rho}
%$$ 
%where $C_V=(3/2){\cal R} (1/A+X_e)$ is the specific heat 
%at constant volume, ${\cal R}$ is the universal gas constant, 
%$A$ is the mean ion mass in atomic units, $X_e$ is the number of free
%electrons per nucleon. 
%The plasma cooling
%function $\epsilon^-$ (erg cm$^{3}$ s$^{-1}$) is almost 
%independent of temperature in the temperature range $2\times 10^6-10^8$ K
%(see above). Assuming the initial temperature $T_0=10^8$K,
%the temperature will linearly decrease with time. 
%The time evolution of 
%ion densities for different chemical compositions are 
%shown in Figures \ref{nt}. 
The calculated spectral lines
are shown in Fig. \ref{sp} (see \cite{Kosenko_et_al_2003} for more detail).   

%\FIG nt ntT86 ntT86me [width=.5\textwidth] Relative number
%densities of ions as a function of the parameter $nt$ for linear
%plasma cooling law (\protect\ref{T(t)}). {\bf Left panel:} solar
%composition; {\bf right panel:} metal-rich plasma (without H and
%He)

\FIGG sp {spz214nh1d1e+012} {spzme214nh1d1e+010}
[width=.5\textwidth] Time evolution of X-ray spectrum (in the
observer frame at the GRB redshift $z=2.14$) for different time
moments. {\bf Upper panel:} solar composition; {\bf Lower panel:}
metal-rich plasma (without H and He).
                                                                                
%figs for time-dep ion species and line spectrum

\section{Time-dependent thermal effects in optical GRB afterglows} 

The GRB explosion in an inhomogeneous medium 
is a generic picture which is responsible 
for the observed variety of effects in X-ray and
optical afterglow light curves
(see \cite{Bisnovatyj-Kogan_and_Timokhin_1997} for pioneer study). We consider two 
possible types of surroundings: a cloudy, patchy medium and
a shell-like structure around the GRB.
%(Fig. \ref{schema}).
The second case could be 
identified with an earlier SN explosion or 
intensive stellar wind of the underlying 
SN-type event. In both cases,  
a fraction of the GRB energy is deposited into the 
surroundings. This energy is thermalized and
radiated away. If plasma of the clouds is optically thin,
it rapidly cools down. The time-dependent
effects from the "blue" (approaching the observer)
gamma-ray beam are expected on the plasma cooling 
timescale $t_c$ after the GRB, while the reverberation effects
could be observed at different wavelengths later on from 
the "red" (receding) jet on timescales $t_2\sim R(t)/c$.
Here $R(t)$ is time-dependent distance of the photosphere (if formed)
in the medium from which low-energy photons are generated.
The maximum time the time-dependent effects from the "red" beam
will be observed is determined by the limiting distance $d$
and also by possible hydrodynamical effects. So the 
evolving "SN-like" features in the 
optical spectra observed in the afterglow of GRB 030329
several days -- a month after the burst (also called SN 2003dh
\cite{Stanek_et_al_2003,Hjorth_et_al_2003}) 
could be a manifestation of such effects and might 
have no direct relation to SN explosion. 

% figs. schema \label{schema}

\begin{figure}[htb]
%\vspace{9pt}
%\framebox[55mm]{\rule[-21mm]{0mm}{43mm}}
\includegraphics[width=\columnwidth]{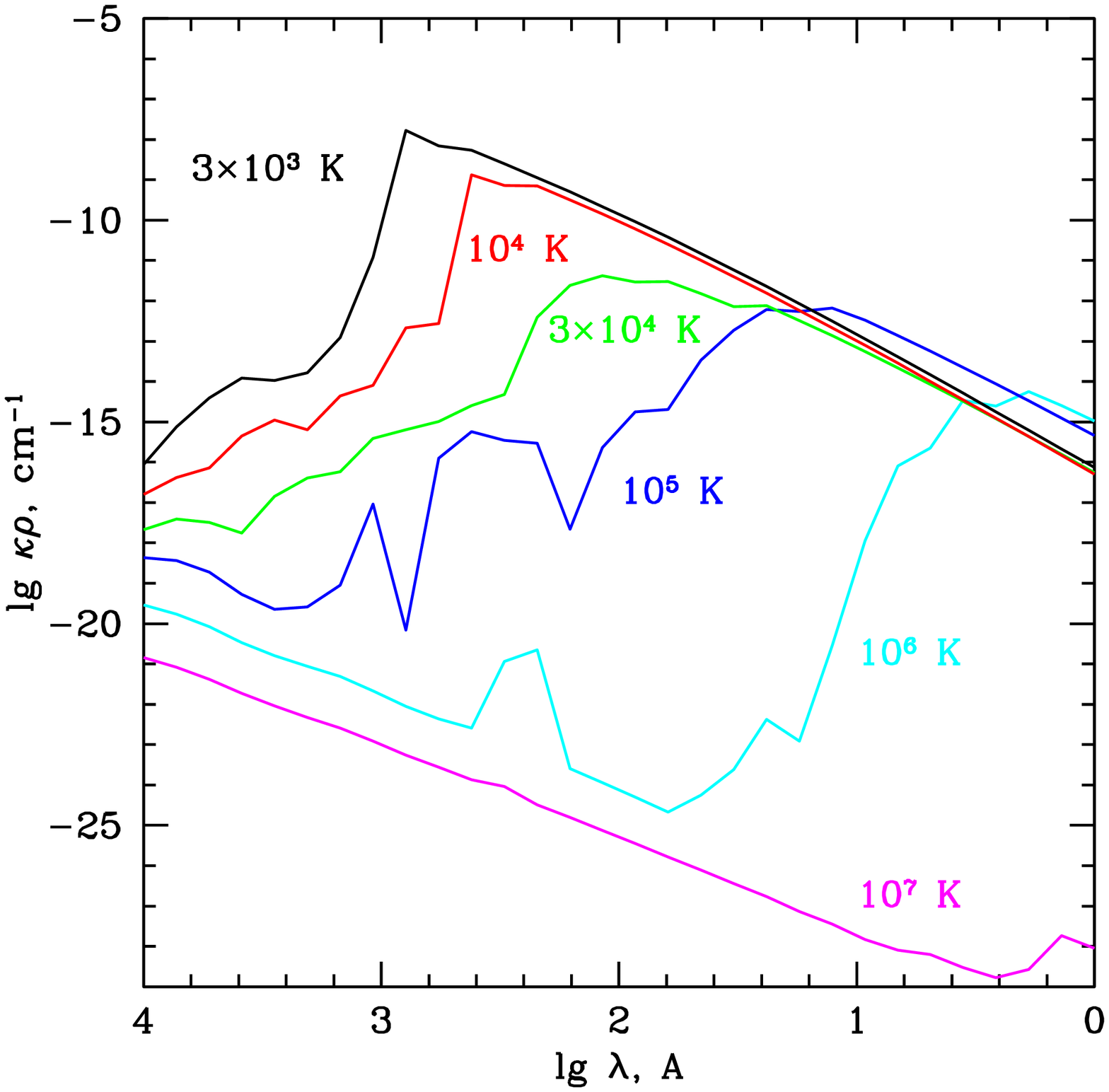}
\caption{Scattering opacity of cooling plasma ($\rho=10^{-14}$ g/cm$^3$) 
with normal abundance.}
\label{opac}
\end{figure}

To illustrate how the cooling plasma forms a photosphere, we 
present the results of LTE calculations of plasma opacities 
as a function of decaying temperature for a fiducial plasma density 
$\rho=10^{-14}$ g/cm$^{3}$ with normal cosmic abundance 
(Fig. \ref{opac}). The scattering 
opacity is $\kappa_T\rho\sim 3\times 10^{-15}$ cm$^{-1}$ and 
is independent of temperature until recombination at low temperatures
lowers it down to $\sim 10^{-19}$ cm$^{-1}$. These calculations 
indicate that up to limiting distances $d\sim 10^{17}$ cm 
(as appropriate to the case of GRB 030329), such a medium 
remains optically thin for scattering, but a photosphere start forming 
for hard photons when temperature drops below $\sim 10^6$ K. 
% figs. opacities \label{opac} 

\begin{figure}[htb]
%\vspace{9pt}
%\framebox[55mm]{\rule[-21mm]{0mm}{43mm}}
\includegraphics[width=\columnwidth]{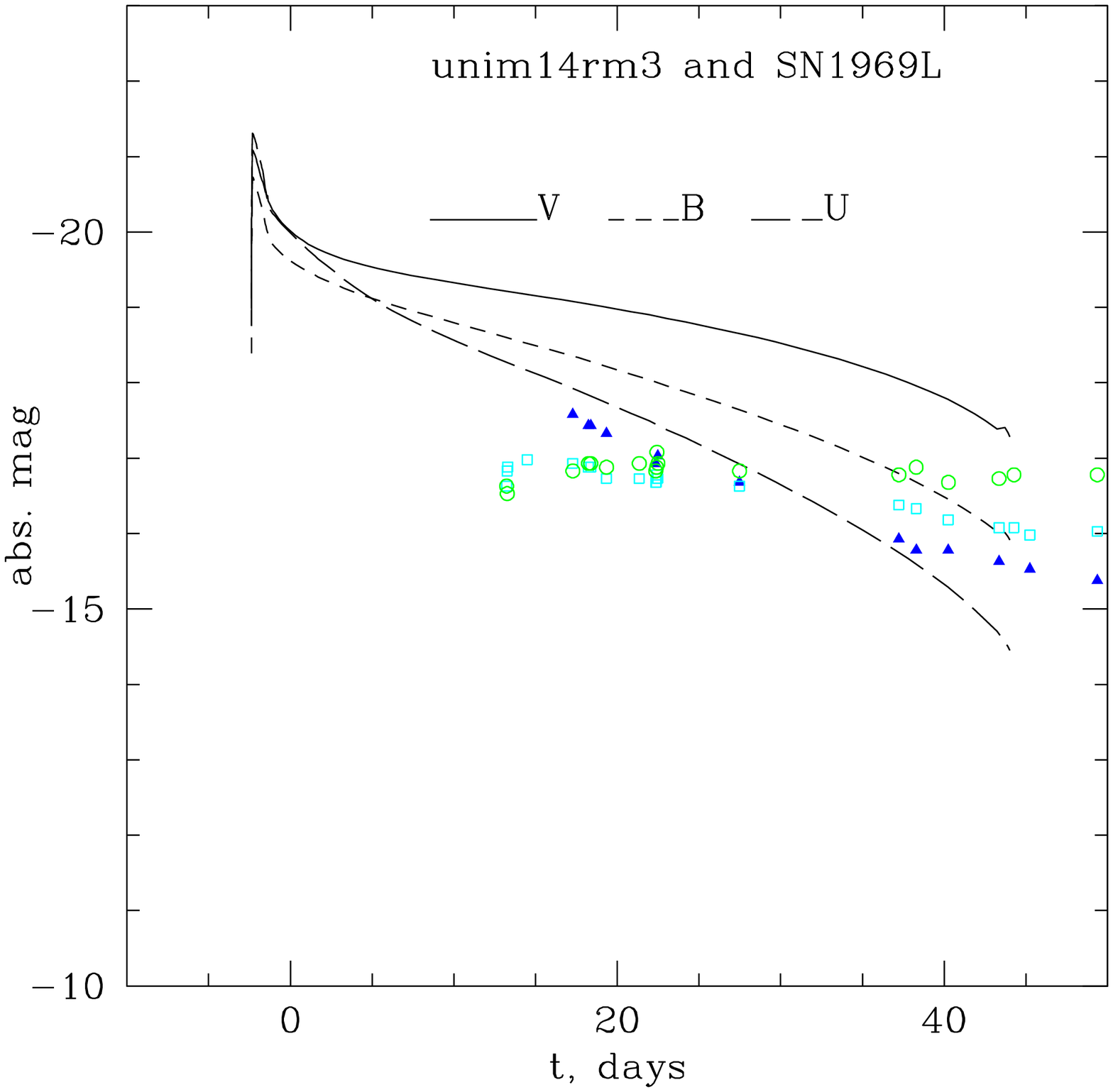}
\caption{The "mini-SN" effect from a large homogeneous cloud
with $R=6\times 10^{15}$ cm after energy deposit 
$\Delta  E \approx 6 \times  10^{49}$ erg. 
SN 1969L is shown for comparison.}
\label{large_cl}
\end{figure}

To have an idea what kind of optical events can be 
expected from clouds illuminated, we apply the multi-group radiation 
hydrocode STELLA \cite{STELLA} 
to calculate UBV-light curves (in the GRB rest-frame).
We assume an energy deposit $\Delta E$ thermalized in 
a homogeneous cloud ($\rho=10^{-14}$ g/cm$^3$) 
of mass $M$ over the GRB duration 
time $\Delta t$. No equilibrium between photons and plasma is assumed.   
These calculations are similar to our "mini-supernova" model \cite{miniSN},
but the difference is that here the cloud is transparent for photons
so the energy is thermalized within the entire volume of the cloud.  
Fig. \ref{large_cl} shows the light curve for a large ($R=6\times 10^{15}$
cm) cloud with a total mass of $\sim 5 M_\odot$, 
$\Delta  E \approx 6 \times 10^{49}$ erg, 
and $\delta t=100 $s. 
The cloud is heated up to 
a temperature of $10^4$ K. The optical 
emission from the cooling cloud continues for more than a month. 
For comparison, U (filled triangles), B (open squares),  V (open circles)
data of a typical  plateau type II SN 1969L are shown.      
A very bright optical outburst (the peak absolute magnitude 
$M_V\sim -21.5^m$)
is expected shortly after the
energy deposit (the light travel time correction
will somewhat smear the peak on the time scale $\sim R/c$.) 
This effect can be compared with the enigmatic outburst 
seen in the R afterglow of GRB 030329 $\sim 1.7$ day after the GRB
\cite{Matheson_et_al_2003} .

\begin{figure}[htb]
%\vspace{9pt}
%\framebox[55mm]{\rule[-21mm]{0mm}{43mm}}
\includegraphics[width=\columnwidth]{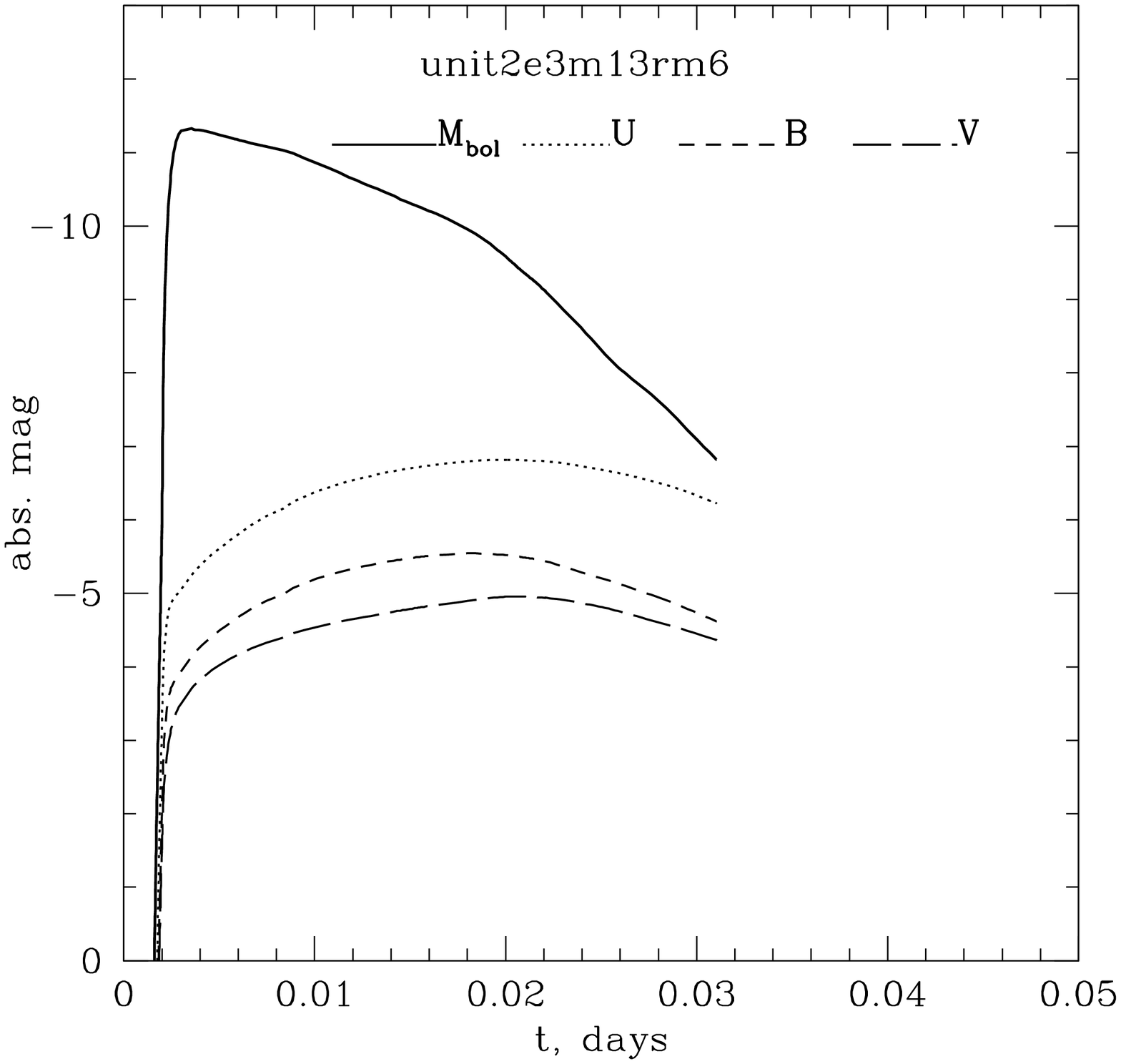}
\caption{Thermal relaxation of a small homogeneous cloud
with $R=1.4\times 10^{13}$ cm after thermalizing $\Delta  E = 10^{43}$ erg
inside the inner 
$\Delta M=6\times 10^{-7} M_\odot$.}
\label{mini_cl}
\end{figure}

A different picture is obtained if we consider a smaller homogeneous 
cloud with $R=1.4\times 10^{13}$ cm and $\rho=10^{-13}$ g/cm$^3$
(i.e. like those required to explain 
non-stationary X-ray emission lines in GRB 011211). An energy 
deposit of $10^{43}$ erg into the  
inner $\Delta M=6\times 10^{-7} M_\odot$ 
over $\Delta t=10$ s heats up the cloud to $T\sim 40000$ K.
This yields (Fig. \ref{mini_cl}) $M_{opt}\sim -5^m...-7^m$.
A collection ($10^5-10^6$) 
of such clouds (patchy presupernova wind?) at $d\sim 10^{17}$
cm would naturally produce a reverberating optical bump with $M_{op}\sim 
-17^m...-19^m$ on the 
optical light curve of the canonical power-law GRB afterglow 
several weeks after the GRB. It is still to be explored if such 
clouds can mimic spectrum of a SN Ic event spectroscopically 
identified in 
GRB 030329 \cite{Stanek_et_al_2003,Hjorth_et_al_2003}.

\section{Conclusions}

Time-dependent thermal plasma effects from GRB surroundings  
should manifest in X-ray and lower energy afterglows of a GRB 
twofold: 

(1) Shortly after GRB (on the plasma cooling time scale determined by
density, chemical composition and the energy deposited over the GRB
emission time) thermal
effects may be produced by plasma heated up by the "blue"
GRB jet. The fading X-ray emission lines in GRB 011211 may
provide an example.
%(1) Shortly after the GRB, on the plasma cooling timescale $t_c$
%which is mainly determined by the surrounding 
%plasma density and chemical composition
%and the energy per nucleon deposited over the GRB emission time.
%They are produced by plasma heated up by the "blue" GRB jet. 
%The fading X-ray emission lines in GRB 011211 provides an example. 
%The model requires a cloudy surroundings with $N_{cl}\sim 10^6$,
%$l_{cl}\sim 10^{12}$ cm, $n_e\sim 10^{11}-10^{12}$ cm$^{-3}$, 
%up to $d\sim 0.1$ pc from  the GRB site. Such a medium can possibly
%exist in star-forming high-redshift GRB host galaxies. The 
%parameters of the clouds are similar to those required by the GRB external
%shock model \cite{Dermer_and_Mitman_1999}. The model allows 
%the independent estimation of the GRB opening angle ($2\theta\sim 0.1$
%for GRB 011211). 

(2) At later times, variable thermal optical features can be 
observed superimposed on the underlying non-thermal afterglow continuum due to 
the contribution from cooling clouds/shells heated by the receding ("red") GRB jet
on timescales limited by the most remote clouds $d$. 
Optical afterglows of GRB 021004 and GRB 030329 deviating from 
pure power-law can provide the example. The effective
photosphere moving across the heated region can mimic the SN-like
effects observed in some GRB.

\end{document}